\documentclass[aps,prb,twocolumn,showpacs]{revtex4}
\usepackage{amsmath}
\usepackage{graphicx}
\usepackage{subfigure}
\begin{document}

\title{Birth, Growth and Death of an Antivortex during the
Propagation of a Transverse Domain Wall in Magnetic Nanostrips}

\author{H. Y. Yuan, X. R. Wang}
\email{[Corresponding author:]phxwan@ust.hk}
\affiliation{Physics Department, The Hong Kong University of
Science and Technology, Clear Water Bay, Kowloon, Hong Kong}
\affiliation{HKUST Shenzhen Research Institute, Shenzhen 518057, China}
\date{\today}

\begin{abstract}
Antivortex birth, growth and death due to the propagation of a transverse
domain wall (DW) in magnetic nanostrips are observed and analyzed.
Antivortex formation is an intrinsic process of a
strawberry-like transverse DW originated from magnetostatic interaction.
Under an external magnetic field, DW in a wider width region
tends to move faster than that of a narrower part.
This speed mismatch tilts and elongates DW centre line.
An antivortex is periodically born near the tail of the DW centre line.
The antivortex either moves along the centre line and dies on the
other side of the strip, or grows to its maximum size, detaches itself
from the DW, and vanishes eventually. The former route reverses the
polarity of DW while the later keeps the DW polarity unchanged.
The evolution of the DW structures is analyzed using
winding numbers assigned to each topological defects.
The phase diagram in the field-width plane is obtained and discussed.
\end{abstract}

\maketitle

\section{Introduction}
Domain wall motion in ferromagnetic nanostrips has attracted much
attention in recent years due to its potential applications in magnetic
memory devices \cite{Parkin1} and magnetic logic gates \cite{Allwood}.
These applications require a detailed and deep understanding of
DW motion so that one could have a better control of magnetization states.
Although field driven DW motion have been extensively studied
\cite{Walker,Wieser,wang1,Beach,Porter,Erskine,wang2,Linder,Parkin2}
in the last few decades and lots of interesting phenomena have been
found, there are still many fundamental processes poorly understood.
For example, field driven DW dynamics is governed by the
Landau-Lifshitz-Gilbert (LLG) equation which has a well-known Walker
exact rigid-body propagation solution \cite{Walker} for a one-dimensional
(1D) biaxial wire. However, simulations or experiments on the field
driven DW motion are often not described by the Walker solution
\cite{Wieser,wang2}. For quasi-1D nanostrips with narrow width,
transverse DW is a preferred low energy state \cite{Porter} as shown in
Figs. 1a and 1b without (1a) and with (1b) the magnetostatic interaction.
The magnetostatic field plays such an important role in the DW
structure that the width of DW varies in the transverse direction,
leading to a strawberry-like DW (Fig. 1b).
Although similar shaped DW has already been known numerically
\cite{Parkin2,McMichael,Nakatani1,Laurson}, its consequences and
connections with other DW dynamics phenomena such as antivortex
generation and reduction of the Walker breakdown field have not been
thoroughly investigated, and this is the main focus of this work.

In this paper we explore the consequences of the strawberry-like
transverse DW on field driven DW dynamics in ferromagnetic nanostrips.
When the field is larger than a certain value but substantially below the
usual Walker breakdown field, a series of DW structure transformations,
including periodic birth and death of antivortices, is observed and explained.
Two types of antivortex evolutions and their causes are identified.
A phase diagram in the field-width plane is also obtained numerically.
This paper is organized as follows. Our model and method is described
in the next section. An explanation of why magnetostatic interaction
should give rise to a strawberry-like transverse DW is also provided
in Section II. Numerical results on the antivortex generation
and its afterward evolution are presented and discussed in Section III.
Discussion and conclusion are given in Section IV, followed by the
acknowledgments.

\section{Model and Method}

Our model is a head-to-head DW in magnetic nanostrips of $5000$nm long,
$4$nm thick, and width varying from $40$nm to $120$nm.
The magnetization dynamics of nanostrips is governed by the LLG equation:
\begin{equation}
\frac{\partial\mathbf{m}}{dt} = - \mathbf{m \times H}_{\hbox{eff}} +
\alpha \mathbf{m} \times \frac{\partial\mathbf{m}}{dt},
\end{equation}
where $\mathbf{m}(\mathbf{x},t)= \mathbf{M}(\mathbf{x},t)/M_s$ is the
unit vector of magnetization $\mathbf{M}$, $M_s\equiv|\mathbf{M}|$ is
the saturation magnetization, $\alpha$ is the phenomenological Gilbert
damping constant. The time $t$ is in the units of $(|\gamma| M_s)^{-1}$
where $\gamma$ is gyromagnetic constant. $\mathbf{H}_{\hbox{eff}} =
\frac{2A}{\mu_0M_s^2}\nabla^2 \mathbf{m}+\mathbf{H}_K +\mathbf{H}_d +
\mathbf{H}_0$ is the effective field measured in the units of $M_s$,
which includes the exchange field described by the exchange constant $A$, 
crystalline anisotropy field $\mathbf{H}_K$, magnetostatic field 
$\mathbf{H}_d$ and applied field $\mathbf{H}_0$. 

The film lies in $xy$plane with the origin at the center of the strip.
$x$ is along the longitudinal direction with $y$ in the width 
(transverse) direction and $z$-axis along the thickness direction.
Permalloy parameters are used in the simulations with 
$A =13 \times 10^{-12}$J/m, saturation magnetization $M_s = 800
\times 10^{3}$A/m and negligible crystalline anisotropy field. Damping
constant is chosen to be $\alpha = 0.1$ to accelerate the simulation.
The LLG equation is numerically solved by the micromagnetic package 
OOMMF \cite{oommf}. In our simulations, the mesh size is chosen to be
4nm$\times$4nm$\times$4nm where 4nm is about 0.7 of exchange length 
$l_{ex}=\sqrt{2A/\mu_0M_s^2}$ \cite{Nakatani1}.

\begin{figure}
\includegraphics[width=8cm]{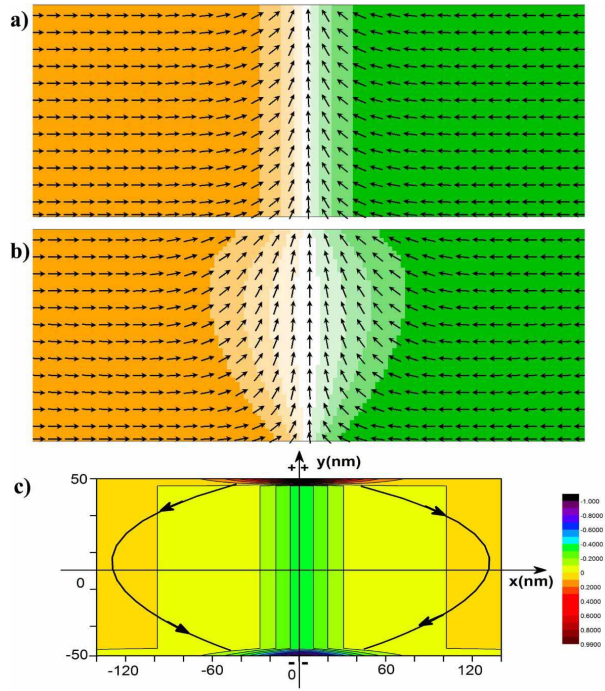}\\
\caption{(Color online) DW structures and magnetic charge distribution
near the DW. For clarity, each spin represents average magnetization of
4 cells of 8nm$\times$8nm$\times$4nm. a) Homogeneous transverse DW when
the magnetostatic interaction due to the edge magnetic charges of DW
is neglected. b) The strawberry-like transverse DW when all
magnetostatic interactions are included. The color codes the
values of $m_x$, varying from orange for $m_x =1$ to green for
$m_x=-1$ with white for $m_x=0$ (DW center).
Both strips have dimensions $3000$nm $\times 100$nm $\times 4$nm.
c) The magnetic charge distribution for DW structure shown in a).
The black curved lines with arrows indicate the magnetic field
generating by the edge charges.}
\label{fig1}
\end{figure}

The static DW is transverse \cite{Porter} for the types of strips in 
the current investigations, as illustrated in Fig. \ref{fig1} which is 
numerically obtained from an initial configuration of $m_x=-\tanh(x/
\Delta)$, $m_y=1/\cosh(x/\Delta)$ and $m_z=0$ for a head-to-head DW, and 
parameter $\Delta$ is close to the static DW width. When the magnetostatic 
interaction includes only the shape anisotropy, but neglects all 
contributions from the non-uniform magnetization distribution, the final 
DW is homogeneous in which DW width is $y-$independent and the net magnetic 
moments point up (polarity $+1$ \cite{Van}) as shown in Fig. \ref{fig1}a. 
When all the magnetic dipole-dipole interaction is included in the 
magnetostatic interaction, the magnetostatic field generated by the 
magnetic charges along the edges (Fig. \ref{fig1}c) shall modify the
effective magnetic anisotropy such that a typical DW shape is strawberry-like
as shown in Fig. \ref{fig1}b. Specifically, the $y-$component 
of edge charges field tends to reduce the anisotropy field in the 
direction while its $x-$component lowers the anisotropy field near upper 
edge and enhances the $x-$component of the field near the lower edge.
Because the decrease of anisotropy leads to the increase of DW width,
this anisotropy modification gives rise to the strawberry-like DW that
closely follow the field lines, as shown in Fig. \ref{fig1}c.
The strawberry-like DW shall be upside down for a transverse wall
with net magnetic moments pointing down (polarity $-1$).
Similar DW shape have been reported before \cite{Parkin2,McMichael,
Nakatani1,Laurson} with a complicated explanation.

It is known that field-driven DW propagating speed is proportional to
the DW width \cite{Walker,wang1}. Thus, under an external field, one
expects that the propagation of a homogeneous DW as shown in Fig.
\ref{fig1}a is described well by the Walker famous solution \cite{Walker}.
For an inhomogeneous DW like that in Fig. \ref{fig1}b, things become 
very different. For example, DW in a wider region tends to move faster 
than that in a narrower one. As a result, the DW centre line must tilt 
and stretch due to the speed mismatch and the exchange interaction,
creating a large magnetization gradient near the tail of the centre line.
Complicated changes could happen such as antivortex generation and 
reduction of the Walker breakdown field. The detail simulation results on
the field-driven strawberry-like transverse DW are presented below with
a qualitative understanding.

\section{Results}

We use a magnetic strip of width $80$nm as an example to illustrate
possible types of field-driven transverse wall motion. Since the only
anisotropy comes from sample shape in the model, the Walker breakdown field
is $H_W=\frac{1}{2}\alpha(N_z-N_y)M_s=417$Oe \cite{Walker,McMichael,Bryan},
where $N_z$, $N_y$ are demagnetization factors along $z-$ and $y-$directions.
For fields below $237$Oe, the transverse wall propagates eventually like a
rigid-body after a short transient process. When the field is greater than
$237$Oe, antivortex is generated and no rigid-body DW propagation is observed.
The possible types of the antivortex evolution depend on field strength.
For fields $237$Oe$\leq H < 255$Oe, antivortices generated at one edge
defect move along its DW centre line to the other edge and die there.
At the same time the polarity of the transverse wall is reversed.

\begin{figure}
\includegraphics[width=8cm]{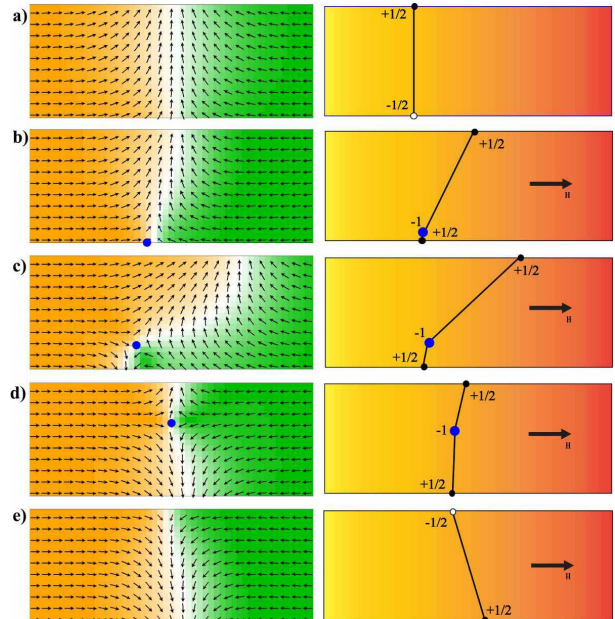}\\
\caption{(Color online) The snapshots of spin configurations (left
plots) and corresponding illustrations (right plots) of topological
defects, big blue dots for antivortices and open circles for edge
defects of winding number $-1/2$ and filled black circles for edge
defects of winding number $1/2$, and DW centre lines.
The width of the strip is 80nm and the applied field is 245Oe.
The initial (before the field is applied) DW structure has polarity $+1$.
$x_t$ is $x-$coordinate of DW centre at the top edge.
a) The initial strawberry-like transverse wall with $x_t$=0nm at $t=0$ns.
b) DW structure with $x_t=525$nm at $t=0.78$ns.
c) DW structure with $x_t=743$nm at $t=1.17$ns.
d) DW structure with $x_t=733$nm at $t=1.66$ns.
e) DW structure with $x_t=898$nm at $t=1.98$ns. }
\label{fig2}
\end{figure}

Fig. \ref{fig2} shows DW transformations under a $245$Oe field.
As shown in the left plot of Fig. \ref{fig2}a, the DW is strawberry-like
before the field is applied. The spins along the DW centre line are
pointing up ($+y$ direction), and DW polarity is defined as $+1$ \cite{Van}.
The DW composes two edge defects whose winding numbers \cite{Chern}
are $1/2$ (filled circle) and $-1/2$ (open circle) on the top and bottom
edges, respectively. The DW centre line is parallel to $y-$axis as illustrated
by the black vertical line in the right plot of Fig. \ref{fig2}a. After a
$245$Oe is applied along $+x$-direction, the centre line tilts its direction
away from $y-$axis and elongates at the same time (Fig. \ref{fig2}b-e).
As the centre line tilts far enough from the $y-$axis as shown in the left
plot of Fig. \ref{fig2}b at $0.78$ns after field is turned on, a small 
antivortex (blue dot) is born at the edge defect on the bottom edge, the 
tail of DW centre line. Since the defects are topological objects with 
well-defined winding numbers \cite{Chern}, edge defect of $-1/2$ winding 
number can only give a birth to an antivortex of winding number $-1$ 
and change its own winding number to a $+1/2$ edge defect\cite{Chaikin}.
This is illustrated in the right plot of Fig. \ref{fig2}b with the blue
big dot representing the antivortex and two filled circles for the edge
defects of $+1/2$ winding number. Then, this antivortex moves toward the
other side of the strip (Fig. \ref{fig2}c at 1.17ns and Fig. \ref{fig2}d
at 1.66ns) along the DW centre line. At 1.98ns, the antivortex reaches
the top edge defect and die there as shown in Fig. \ref{fig2}e.
At the same time, top edge defect changes its winding number from $+1/2$
to $-1/2$, and transverse wall reverses its polarity (from $+1$ to $-1$).
After a transient period, the above mentioned process will repeat again,
starting from the top edge of the strip this time. This birth-death 
process of antivortices repeats periodically while the transverse DW 
propagates along the while with an averaged speed of about $440$m/s.
This result is also consistent with previous studies
\cite{Beach,Nakatani2,Cowburn,SKK,Kunz,Zinoni,Thiele,Glathe}.

\begin{figure}
\includegraphics[width=8cm]{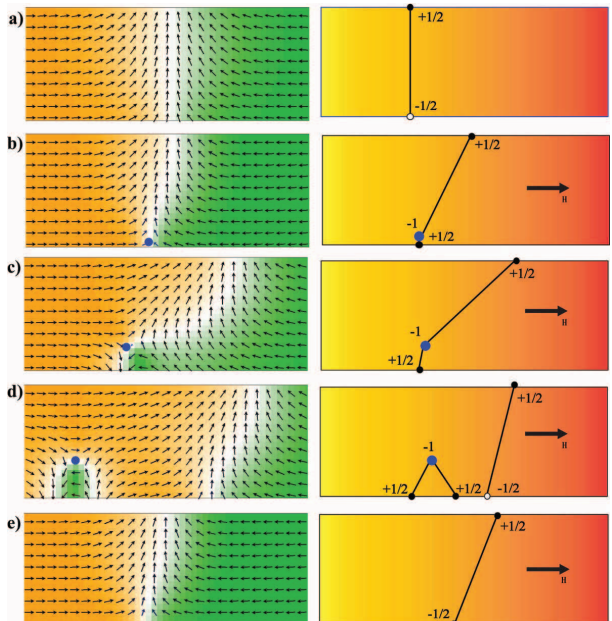}\\
\caption{(Color online) Similar to Fig. 2 but with an applied field of
260Oe. Again $x_t$ is x-coordinate of DW centre at the top edge.
The averaged DW speed is about 587m/s.
a) The initial strawberry-like transverse wall at $t=0$ns with $x_t=0$nm.
b) DW structure with $x_t=365$nm at $t=0.54$ns.
c) DW structure with $x_t=486$nm at $t=0.76$ns.
d) DW structure with $x_t=599$nm at $t=1.02$ns.
e) DW structure with $x_t=769$nm at $t=1.27$ns. }
\label{fig3}
\end{figure}

For a field greater than $255$Oe, an antivortex will not be able to
reach the other edge of the strip because distortion of the DW centre
line is too large so that the corner space is not large enough to 
accommodate the antivortex. Fig. \ref{fig3} is DW transformations under a 
$260$Oe field. The left plot of Fig. \ref{fig3}a shows the initial spin 
configuration of the strawberry-like transverse wall of polarity $+1$. 
The edge defects with corresponding winding number $\pm 1/2$ and the DW 
centre line is illustrated in the right plot of Fig. \ref{fig3}a. 
The internal DW structure at 0.54ns, right before the generation of an 
antivortex at the edge defect on the bottom edge, is shown in the left 
plot of Fig. \ref{fig3}b. The right plot of Fig. 
\ref{fig3}b are the corresponding DW centre line and antivortex (blue 
dot) and two edge defects of winding number $1/2$. Different from the 
previous case, this newly born antivortex cannot reach the other edge. 
The left plot of Fig. \ref{fig3}c is the DW configuration at 0.76ns. 
The corresponding location of the antivortex, edge defects, as well as
the DW centre line are indicated in the right plot of Fig. \ref{fig3}c.
The antivortex makes a U-turn back to bottom edge. Together with two edge
defects of winding number $1/2$ at the bottom edge, the antivortex detaches
itself from the DW by creating an isolated defect region, leaving the DW
to be transverse with polarity $1$ as shown in Fig. \ref{fig3}d at 1.02ns.
The isolated defect region is topologically trivial, and vanishes through
dissipating its extra energy. As shown in Fig. \ref{fig3}e at 1.27ns,
the DW keeps its polarity unchanged at the end of this process, and this
birth-death process of an antivortex repeats itself periodically.
Furthermore, to numerically demonstrate the above antivortex generation 
originates from the magnetostatic field of DW structure, we also carried 
out the OOMMF simulation without this local magnetic charge interaction. 
No antivortex generation is obtained below the Walker breakdown field, 
and the transverse wall always ends up with a rigid-body propagation as 
what was predicted by the Walker solution.

\begin{figure}
\centering
\includegraphics[width=8cm]{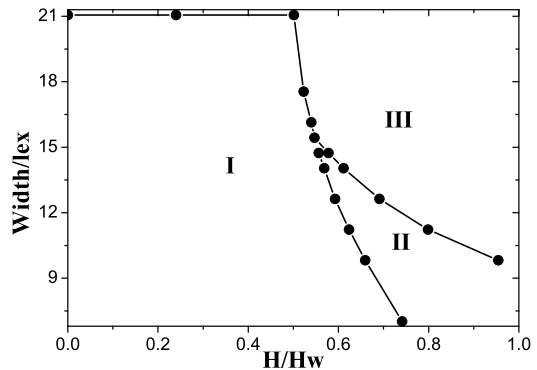}\\
\caption{Phase diagram of three possible phases in the plane of field and
strip width. The field is in units of $H_W$ and width in units of $l_{ex}$.
Phase I is the rigid-body propagation mode. Phases II and III denote
antivortex formation modes. Phase II corresponds to case that an
antivortex can move from one side of strip to the opposite side,
and at the same time the DW polarity is reversed. Phase III is the case
that an antivortex is born, grows to its maximal size and
detaches itself from the transverse wall, then eventually dies inside the
strip. In phase III, DW polarity does not change. }
\label{fig4}
\end{figure}

In order to obtain the phase diagram of these two types of antivortex
evolution in the parameter space of field and strip width, we repeat
the same OOMMF simulations for various strip width. Fig. \ref{fig4} is the
numerical results. Below the width $15.44$ (in units of $l_{ex}$), the
Walker rigid-body DW propagation (phase I) exists at low fields.
At the intermediate fields, a propagation of a strawberry-like transverse
wall gives birth to an antivortex that moves to the other edge and
disappears there. At the same time, the transverse wall changes its polarity
to the opposite sign. This is denoted as the phase II in Fig. \ref{fig4}.
In the phase III, the generated antivortex cannot move to the other edge.
It grows to its maximal size and detaches itself form the transverse wall
by forming an isolated defect region that disappears eventually through
energy dissipation. Above the width $15.44$, we don't observe phase II within
the field step of $1$Oe and only phase I and phase III are observed.

\section{Discussion and Conclusion}

Our study shows that the birth of antivortices arises from the motion of a 
strawberry-like DW caused by the anisotropy modification due to the magnetic
charges of the DW. The uneven energy dissipation in the transverse direction 
due to the DW width variation leads to an uneven propagation speed along 
the direction. As a result, the DW centre line is elongated and tilted.
A large magnetization gradient near the tail of the DW centre line is then 
produced and this gradient tends to generate antivortices near the tail of the 
line. It is clear that the Walker rigid-body propagating mode applies only 
to a transverse DW (Bloch or Neel type), not for an vortex or antivortex DW.
This is because a vortex tends to move in the transverse direction under a 
gyrotropic force \cite{Thiele}. A transversally moving vortex wall in a 
nanostrip must modify its own DW structure due to the edge charge effects.
Thus, DW propagating speed shall vary with time according to the connection 
between DW structure and instantaneous DW speed \cite{wang1}.
Thus, the antivortex birth invalidates the Walker rigid-body propagating mode.
This is why the Walker breakdown field is substantially reduced in a strip.
The actual amount of breakdown field reduction depends on the geometry and 
other material parameters. In general, the larger the strip width is, the 
greater reduction of the Walker breakdown field will be. 

In conclusion, we investigate field-driven DW propagation along magnetic
strips below the predicted Walker breakdown field. A strawberry-like DW
causes reduction of Walker breakdown field because of the antivortices 
generation. The Walker rigid-body DW propagation mode is only possible 
at weak fields when a propagating DW does not create antivortices. 
If external fields are larger than a critical value, this strawberry-like 
DW generates antivortices at the tail of the DW centre line, and the 
antivortices move to the other side of strip and reverse the DW polarity. 
This process prefers to appear in narrower strips according to our simulation. 
If the field is further increased above another critical value, still below 
the Walker breakdown field, the generated antivortex will not be able to reach 
the other edge of the strip, but detaches itself from the DW and vanishes
eventually in a domain through energy dissipation.  The original
transverse DW does not change its polarity in this field range.
It is noticed that motion of an inhomogeneous DW generates antivortices 
even in zero damping case whose details need further investigations. 
\begin{acknowledgments}
HYY acknowledges the support of Hong Kong PhD Fellowship. He would also
like to thank Xiansi Wang, Bin Hu, Yin Zhang and Chen Wang for helpful
discussions. This work was supported by Hong Kong RGC Grants
(604109 and 605413), and the grant from NSF of China.
\end{acknowledgments}


\end{document}